\documentclass[12pt]{article}
\usepackage{a4}
\usepackage{epsf}
\usepackage{amssymb}
\newcommand{\beq}{\begin{equation}}
\newcommand{\eeq}{\end{equation}}
\newcommand{\bea}{\begin{eqnarray}}
\newcommand{\eea}{\end{eqnarray}}
\begin{document}
%\date{\today}

\title{\bf Asymptotic scaling of the gluon \\ propagator on the lattice}
\author{ D. Becirevic$^a$, Ph. Boucaud$^b$,  J.P. Leroy$^b$, 
J. Micheli$^b$, \\  O. P\`ene$^b$, J. Rodr\'\i guez--Quintero$^b$  
and C. Roiesnel$^c$ } \par \maketitle
\begin{center}
$^a${\sl INFN, Sezione di Roma, P.le Aldo Moro 2, I-00185 Rome, Italy} \\
$^b${\sl Laboratoire de Physique Th\'eorique~\footnote{Unit\'e Mixte de Recherche
 - UMR 8627}\\
Universit\'e de Paris XI, B\^atiment 211, 91405 Orsay Cedex,
France}\\$^c$ {\sl Centre de Physique Th\'eorique\footnote{
Unit\'e Mixte de Recherche C7644 du Centre National de 
la Recherche Scientifique\\ 
\\{\tt e-mail: Philippe.Boucaud@th.u-psud.fr, roiesnel@cpht.polytechnique.fr
}}de l'\'Ecole Polytechnique\\
91128 Palaiseau Cedex, France }

\end{center}

\begin{abstract}
We pursue the study of the high energy behaviour of the gluon propagator on the lattice 
in the Landau gauge in the flavorless case ($n_f=0$). It was shown in a preceding paper that the gluon propagator
did not reach three-loop asymptotic scaling at an energy scale as high as 5 GeV. 
Our present high statistics analysis includes also a simulation at $\beta=6.8$
($a\simeq 0.03$ fm), which allows to reach $\mu \simeq 10$~GeV. Special care 
has been devoted to the finite 
lattice-spacing artifacts as well as to the finite volume effects, the latter
being acute at $\beta=6.8$ where the volume is bounded by technical limits.  
Our main conclusion is a strong evidence that the gluon propagator
 has reached three-loop asymptotic scaling, 
at $\mu$ ranging from 5.6~GeV to 9.5~GeV. We buttress up this conclusion
 on several demanding criteria of asymptoticity, including scheme independence.
Our fit in the  5.6~GeV to 9.5~GeV window yields 
$\Lambda^{\overline{{\rm MS}}} \ = \ 319 \pm 14 \ ^{+10}_{-20}$ 
MeV, in good agreement with our previous result, 
$\Lambda^{\overline{{\rm MS}}} \ = \ 295 \pm 20 
$ MeV, obtained from the three gluon vertex, but it is significantly above 
the Schr\"odinger functional
method estimate : $238 \pm 19$ MeV. The latter difference is not understood.
Confirming our previous paper, we show that a fourth loop is necessary to
fit the whole ($2.8 \div 9.5$)~GeV energy window.
 
\end{abstract}
\begin{flushleft}
%\preprint{LPT Orsay-99/nn} 
LPT Orsay/99-72
\end{flushleft}
\newpage
\setcounter{footnote}{0} 
In previous works, we tackled the non-perturbative calculation of the QCD running coupling constant in 
two different ways: (i) by using the three gluon coupling~\cite{alpha,alles}, and (ii) by  matching 
the behaviour  of the lattice regularized gluon propagator to the one predicted by perturbation 
theory~\cite{propag}. The latter method was expected to take benefit of the very good statistical accuracy 
of the propagators and thus to yield a rather precise estimate of the strong coupling constant in its 
ultraviolet (UV) regime, {\it i.e.} of $\Lambda_{\rm QCD}$. Unluckily we
could not fulfill this program for the unexpected reason that  {\sf the gluon propagator has not yet 
reached the asymptotic scaling at scales of $\mathsf{ 2.5 \div 5.0}$ GeV}. This conclusion was supported 
by several different tests. In particular, the remainder of a strong scheme dependence when using one-, two- 
and three-loop formulae indicated the compelling need of higher orders in the perturbative expansion.
 Still we observed that the inclusion of 
third-loop corrections improved the asymptoticity (albeit not enough) over the two-loop results. 
 We were naturally tempted 
to extend the analysis to higher energy scales where any perturbative expansion, with a fixed
number of terms, should progressively improve. This is the basic 
motivation of the present paper.

Since we want to reach ever larger momenta on the lattice, we have to assure that the dominant lattice 
artifacts are under control. We must also ensure that the energy window, in which we could test 
scaling of the gluon propagator, is large enough. The reason is that at large scales, the coupling 
constant has a very mild logarithmic dependence. If the window is too narrow, the higher order terms 
might mimic the lower order ones, that we consider, and thus introduce a bias in $\Lambda_{\rm QCD}$. A 
very wide energy window has been explored in ref.~\cite{luscher} by using the Schr\"odinger functional 
technique. When using the methods based on Green functions~\cite{alpha,alles,propag}, it is more 
difficult to vary the energy scales on several orders of magnitude. Moreover, one deals with more 
scales: the lattice spacing $a$, the linear lattice extension $L$, and the momenta of the gluons 
$p^2$. On the other hand, compared to the  Schr\"odinger functional method, we believe that the 
Green functions have a simpler physical meaning and, being conceptually very different, represent 
a necessary test. 

\par
The requirement of larger momentum scales implies smaller lattice spacings if we are to keep the 
UV ${\cal O}(a^2p^2)$ artifacts under control ($ap\ll 1$). Equivalently, we need to perform the 
simulations at larger $\beta$, which (for reasonable computing time) also means smaller volumes 
and potentially dangerous infrared (IR) finite volume artifacts. To prevent these problems, we need 
to ensure $Lp \gg 1$. 
 
\par
The question is whether we can find an ensemble of lattice results~\footnote{By a {\sl lattice result}, 
we mean a value of the bare propagator for one value of $p^2$ obtained on a lattice with a given $\beta$, 
and in a specific volume.} fulfilling
all requirements, {\it i.e.} that the lattice artifacts are small enough (${\cal O}(L^{-1})\ll p\ll {\cal O}(a^{-1})$), and 
that the energy window
is large. This is a very demanding requirement because a small change in the coupling constant can induce 
a large uncertainty in $\Lambda_{\rm QCD}$. To avoid the statistical uncertainties in that respect, we 
work with $1.000$~independent gauge field configurations at every stage of this research~\footnote{The 
only exception is the simulation at $(6.0, 32^4)$, where we have $100$ configurations.}. To keep 
the energy window large enough, we need to fit simultaneously the lattice data obtained at different 
lattice spacings. In other words, we must ensure that the small momentum lattice data with large $\beta$, 
and large momentum data with small (but reasonable) $\beta$, are compatible. In order to achieve that, 
and to reduce systematic uncertainties to the level of statistical ones, one evidently needs to control 
both IR and UV lattice artifacts for all the data. 
Therefore, a major part of this paper is devoted to these issues: the reduction of the UV and IR 
systematic uncertainties. The discretization errors are monitored by working at three values of the 
bare lattice coupling:
\beq
 \beta \ =\ \left\{  \hspace*{7mm} 6.0,\hspace*{12mm} 6.2, 
 \hspace*{14mm}  6.8 \hspace*{7mm} \right\}\;. 
\eeq
A lattice spacing $a^{-1}=2.72(11)$~GeV has recently been measured at $\beta=6.2$ with 
a non-perturbatively 
improved action~\cite{phi}.  For a direct comparison with~\cite{alpha,alles,propag}, in this
paper we will keep $a^{-1}(\beta=6.2)=2.75$~GeV which is well within the error bars.
Other lattices are calibrated relatively to this one, by using the lattice measurement of the string 
tension~\cite{bali}. We take:
\beq
 a^{-1} \ =\ \left\{ \ 1.97\ {\rm GeV},\ 2.75\ {\rm GeV},\ 6.10\ {\rm GeV} \ \right\} \;,   
\eeq
at $\beta$ = 6.0, 6.2 and 6.8 respectively.
Thus, our lattice spacings vary from $0.03$~fm to $0.10$~fm. The study
at $\beta=6.8$ ($a\simeq 0.03$~fm) allows us to reach momenta up to $\sim 10$~GeV. The main study of the 
finite volume effects is performed at $\beta = 6.0$, by repeating the
calculation with the following 
lattice volumes:
\bea
V\ =\ \left\{ \ 12^4,\ 16^4,\ 24^4,\ 32^4 \ \right\}\; . 
\eea
From this study we deduced an efficient parametrization of the finite volume effects~(\ref{vol}), which 
allows us to extrapolate our high momentum data to the $V\to \infty$ limit. An important cross check is 
provided by two volumes at $\beta = 6.8$:
\bea
V\ =\ \left\{ \ 16^4,\ 24^4 \ \right\}\; , 
\eea
while for $\beta = 6.2$, we work with the volume $V=24^4$ only.

An improved version of the method used in ref.~\cite{propag} to cure lattice hypercubic effects, 
has been applied. The non-hypercubic finite spacing effects have been dealt with by comparing 
different values of $\beta$.

Curing the above-mentioned artifacts, we could 
keep most of our lattice points. We discarded those which exhibit large corrections.

\par
In this paper, we will not deal with the small momentum behaviour
of the gluon propagator. We postpone it to our forthcoming publications. This part has so far attracted 
a lot of attention
in the literature~\cite{bernard}-\cite{ma} (for a review with a rather complete list of references, 
see ref.~\cite{mandula}),
while, to our knowledge, the large momentum gluon propagator has not been studied in detail.
 Preliminary results 
of this paper were presented in ref.~\cite{latt99}.

The remainder of this paper is organized as follows: In section 1 we outline the generalities of the 
method we use (previously described in ref.~\cite{propag}), and introduce the main perturbation
theory tools. In section 2, the lattice artifacts are discussed. Those related to hypercubic geometry 
are eliminated by improving the method presented in ref.~\cite{propag}. A procedure allowing to treat 
empirically the finite-volume effects is described. In section 3 we perform a three-loop fit to the 
gluon propagator in the $\widetilde{\hbox{MOM}}$ scheme and apply the test of scheme independence, 
by considering a
$(5.6\div 9.5)$~GeV window in which only the data at $\beta=6.8$ are used. 
Section 4 is devoted to the study of the whole window, ranging from 2.8 up to 9.5 GeV by combining all 
lattice data. We discuss our results in section 5 and conclude in section 6.

\section{General description of the method}
\label{method}

The Euclidean two point Green function in momentum space writes in the Landau gauge: 
\beq
        G_{\mu_1\mu_2}^{(2)\,a_1 a_2}(p,-p)\ =\ 
        \delta_{a_1 a_2} \left(\delta_{\mu_1\mu_2}-
        \frac{p_{\mu_1}p_{\mu_2}}{p^2}\right)\ G^{(2)}(p^2)\,,
\label{G2}
\eeq
where $a_1, a_2$ are the color indices ranging from 1 to 8. 
The bare gluon propagator in the Landau gauge (see for instance ref.~\cite{propag}), is such that
\beq
\lim_{\Lambda\to \infty}
\frac{d \ln Z_3(\mu,\Lambda)}{d \ln \mu^2} \; \equiv \;
\lim_{\Lambda\to \infty}
\frac{d \ln [\mu^2 G^{(2)}_{\rm bare}(\mu,\Lambda)]}{d \ln \mu^2}\,,
\label{obs}\eeq 

\noindent  is independent of any
regularization scheme. $Z_3(\mu,\Lambda)$ is the gluon renormalization constant in the MOM 
(or $\widetilde{\rm MOM}$) scheme at the point $p^2=\mu^2$, and $\Lambda$ is a generic notation 
for the UV cut-off ($a^{-1}$ or $(d-4)^{-1}$).
 
 It is well known that the $\mu$ and the $\Lambda$
dependences of $Z_3(\mu,\Lambda)$ factorize when one drops 
all the terms vanishing as $\Lambda \to \infty$ (see ref.~\cite{grunberg}), and we can write:
\bea
Z_3(\mu,\Lambda)=Z_3^R(\mu) Z_3^b(\Lambda) + {\cal O}(1/\Lambda) \;\; ;
\label{fact}
\eea

\noindent where the evolution of both, $Z_3^R(\mu)$ and $Z_3^b(\Lambda)$, is described by 
Callan-Symanzik equations

\bea
\left({d \over d \ln \mu^2}- \Gamma^R(\mu)\right) \ Z_3^R(\mu) \ = 0 \ , \nonumber \\
\left({d \over d \ln \Lambda^2}- \Gamma^b(\Lambda)\right) \ Z_3^b(\Lambda) \ = 0 \ .
\label{callan-symanzik}
\eea
From the QCD perturbation theory we know that
\beq
\label{gamma}
\frac{d \ln Z_3^R(\mu)}{d \ln \mu^2}=\Gamma^R(\mu)=-\left( \frac{\gamma_0}
{4\pi} \alpha +\frac{\gamma_1} {(4\pi)^2} \alpha^2
+\frac{\gamma_2} {(4\pi)^3} \alpha^3 + O(\alpha^4)\right)\,,
\eeq
where it is understood that the coupling constant in a given scheme is a function of $\mu$ such that
\beq
\label{beta}
\frac{\partial \alpha}{\partial \ln \mu}=\beta(\alpha)= -\frac{\beta_0}
{2\pi} \alpha^2 -\frac{\beta_1} {(2\pi)^2} \alpha^3
-\frac{\beta_2} {(4\pi)^3} \alpha^4 + O(\alpha^5)
\eeq
with
\beq
\beta_0=11,\quad \beta_1=51,\quad \gamma_0=\frac {13}{2}\,, \label{gamma0}
\eeq
in the flavorless case ($n_f=0$), while $\gamma_1,\gamma_2$ and $\beta_2$ 
are scheme dependent. To be 
specific, in the flavorless $\widetilde{\hbox{MOM}}$  
scheme\footnote{Details on the computation of 
the parameters $\beta_2$, $\gamma_1$, $\gamma_2$ 
in this scheme can be found in refs.~\cite{propag,beta2}.}:
\beq
\beta_2\simeq 4824.,\quad\gamma_1= \frac {29}8,\quad \gamma_2 \simeq 960\,.
\label{beta2}
\eeq

Lattice calculations provide us with the bare propagator but in a finite volume which, besides the UV 
cut-off ($\Lambda \sim 1/a$), introduces an additional length dimension, $L$ (the physical volume 
being $L^4$). As we shall see, finite volume effects, as well as hypercubic artifacts, should be 
eliminated first in order to have access to the renormalization constant in eq.~(\ref{obs}), 
$Z_3(\mu,1/a)$.

Equations formally analogous to (\ref{gamma}) and (\ref{beta}) can be obtained from the second 
line of  eq.~(\ref{callan-symanzik}), with the substitutions of $1/a$ for $\Lambda$ and of the 
lattice bare  coupling constant, $\alpha^b=3/(2\pi \beta)$, for the renormalized one. Unhappily 
the anomalous dimension coefficients, $\gamma_1^b, \gamma_2^b$, have not been determined to our 
knowledge, presumably due to the difficulty of the task. Any perturbative calculation of $Z_3^b(1/a)$ 
appears thus to be limited to one loop, 

\beq
{d \ln Z_3^b(1/a) \over d \alpha^b(1/a)}\ = {\gamma_0^b \over \beta_0} {1\over \alpha^b(1/a)}
\label{Zb}  
\eeq

\noindent for which it can easily be proven that $\gamma_0^b=-\gamma_0$. 

Our general {\sl ``strategy''}, as explained in ref.~\cite{propag}, will be to integrate simultaneously 
eqs.~(\ref{gamma},\ref{beta}), up to three loops in a given scheme.
The solutions depend on the initial values $Z^R_3(\mu_0)$ and      $\alpha(\mu_0)$. They are related 
to the lattice results, $Z_3(\mu,1/a)$, through eq.~(\ref{fact}). When we restrict our computation to 
the lattice data at $\beta=6.8$, $Z^b_3(1/a)$ is just an overall irrelevant multiplicative constant. 
With our three values of $\beta$, two ratios of three ($Z^b_3(6.8)/Z^b_3(6.2)$, $Z^b_3(6.8)/Z^b_3(6.0)$, 
$Z^b_3(6.2)/Z^b_3(6.0)$) are necessary for appropriate matching of all our lattice data. These ratios 
will be fitted and compared to one-loop predictions in section 4. 

Finally, the knowledge of $\alpha(\mu_0)$ in a given scheme allows the determination of $\Lambda_{QCD}$ 
in this scheme, and hence of $\Lambda_{\rm QCD}^{\overline {\rm MS}}$.

\section{Lattice artifacts}

We refer to ref.~\cite{alpha} for technical details concerning the lattice setup in our simulations, 
the calculation of the Green functions, their Fourier transform, the
checks of the $\delta_{a_1,a_2}$ color dependence of the propagators, and the set of momenta 
considered for the different lattices studied. Since the release of ref.~\cite{alpha}, we 
increased the statistical quality of our data, and further explored various lattice volumes 
and various values of $\beta$. As mentioned in the introduction, of special interest for this study 
are the results of our simulation performed at $\beta=6.8$, at two volumes: $16^4$ and $24^4$. 
The high statistical accuracy of our data made a detailed study of systematic uncertainties possible 
and mandatory.

\subsection{Hypercubic artifacts and other $(\lowercase{a}^2 \lowercase{p}^2)$ effects}

We start with the discretization errors. In a finite hypercubic volume the momenta are the discrete sets of 
vectors 
\beq
p_\mu\ =\ \frac{2\pi}L \ n_\mu \;, \label{momentum}
\eeq

\noindent where the components of $n_\mu$ are integers and $L$ is the lattice size. The propagators 
have been averaged as usual, over the hypercubic isometry group $H_4$. The momenta corresponding to 
different orbits of $H_4$, but belonging to the same orbit of the {\sl continuum} isometry group 
$SO(4)$ ({\it e.g.} $n_\mu=(2,0,0,0)$,
and $n_\mu=(1,1,1,1)$), have been analyzed according to an improved version of the method proposed 
in ref.~\cite{propag}.

Let us briefly recall the elements of that method. The main idea is based on the fact that, on the 
lattice, an invariant scalar form factor, like $G^{(2)}(p^2)$, is indeed a function of 4 
invariants $p^{[n]}\equiv \sum_{\mu}p_{\mu}^n$, $n=2,4,6,8$. We will neglect the invariants 
with degree higher than 4 since, in any case, they vanish at least as $\sim a^4$. Thus, we 
parametrize and expand the lattice two-point scalar form factor as a function of the two remaining invariants 
which, on dimensional grounds, appear as $p^2$ and $a^2 p^{[4]}$:
\bea
G_{\rm lat}^{(2)}(p^2,a^2 p^{[4]};L,a) = G_{\rm lat}^{(2)}(p^2,0;L,a) \ + \ \left. 
{\partial G\over \partial (a^2 p^{[4]})} \right|_{a^2 p^{[4]}=0} a^2 p^{[4]} \ .
\label{roies}
\eea
$\left. {\partial G/\partial (a^2  p^{[4]})}
\right|_{a^2 p^{[4]}=0}$ is a symbolic notation for the {\it lattice hypercubic slope}.
This equation summarizes our method to reduce hypercubic 
artifacts and contains all the assumptions on which the method relies. We make the hypothesis that the 
lattice propagator for the discrete momenta belonging to different $H_4$-orbits takes values according 
to a certain
 continuous functional behaviour. When several orbits exist for one $p^2$, to the extent that a linear 
 approximation is licit, the {\sl hypercubic slope} can be extracted and the extrapolation to 
 $a^2p^{[4]}=0$, by using eq.~(\ref{roies}), can be made. We have checked that the linear 
 approximation~(\ref{roies}) is indeed good enough. This is what has been done in ref.~\cite{propag}. 
 Now we elaborate further on this method in order to improve its accuracy and extend its applicability.

\par
\begin{figure}
%\vspace*{-1.cm}
\begin{center}
\begin{tabular}{c c c}
(a)& & \\ 
 &\epsfxsize10.0cm\epsffile{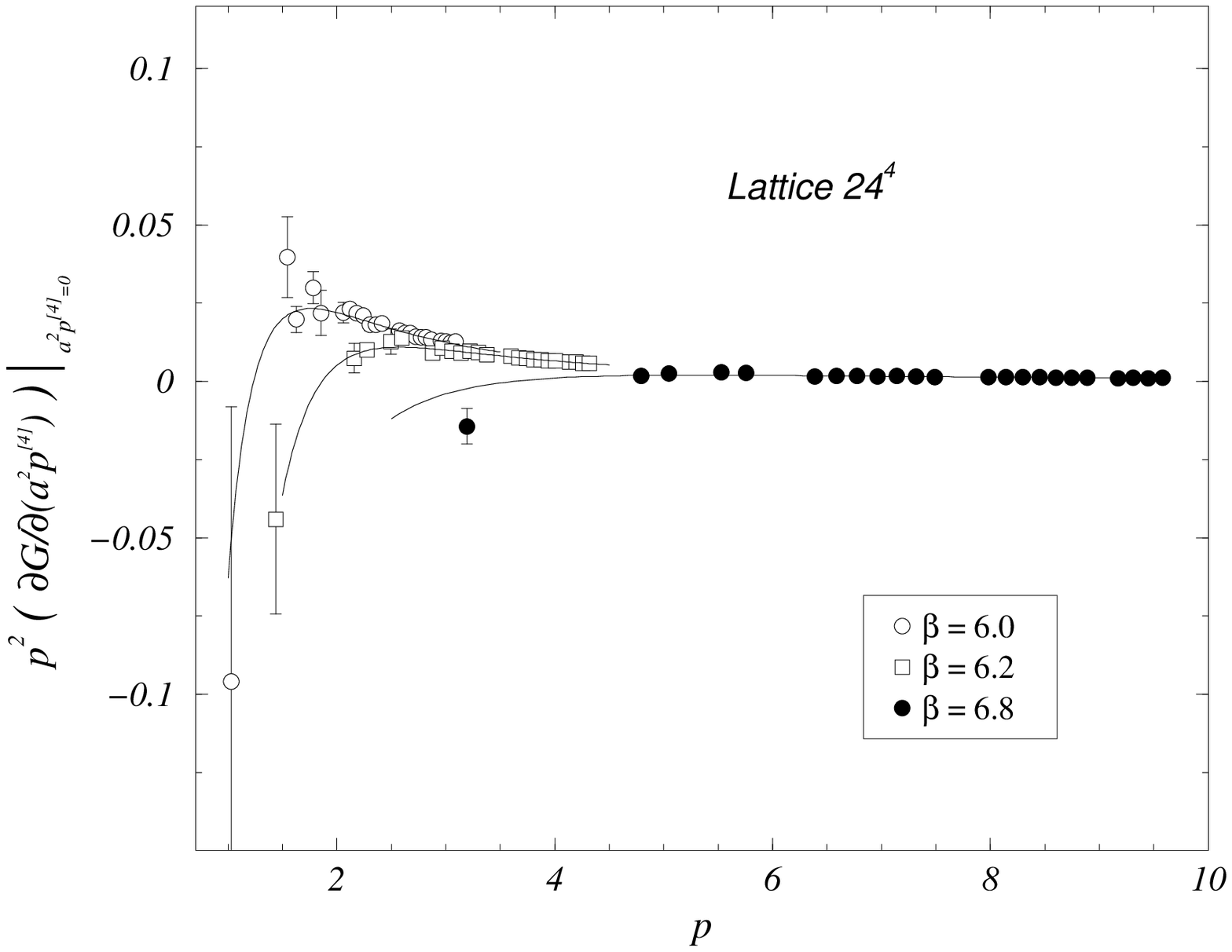} & \\ 
(b)& & \\ 
 &\epsfxsize9.65cm\epsffile{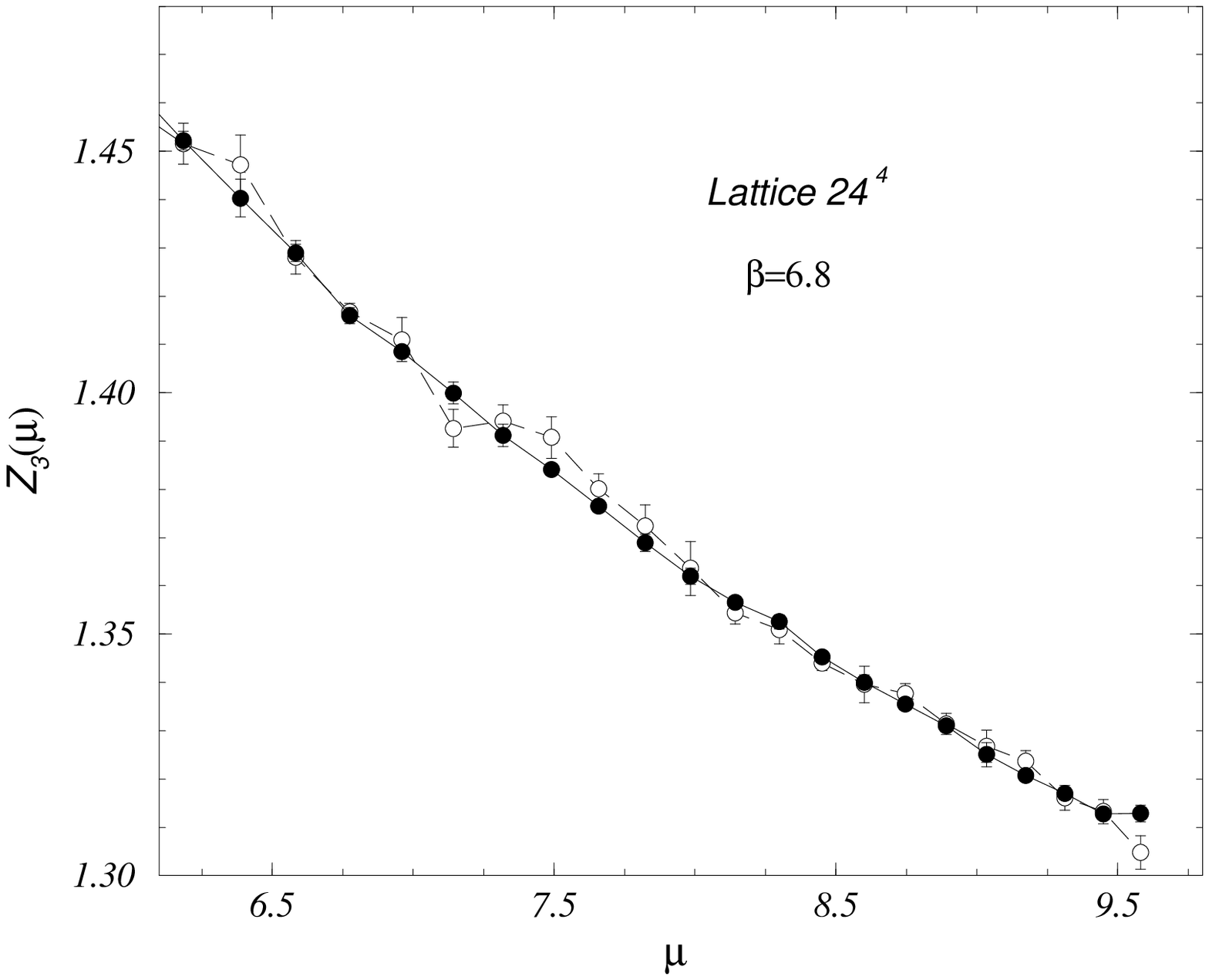} & \\
\end{tabular}
%%%%%%%%%%%%%%%%%%%%%%%%%%%%%%%%%%%%%%%%%%%%%%%%%%%%%%%%%%%%%%%%%%

\caption{\label{figpentes}{\sl 
Plot (a) shows $p^2 \left. {\partial G\over a^2 \partial p^{[4]}} \right|_{a^2 p^{[4]}=0}$, as a 
function of the scale
 $p$  evaluated on $24^4$ lattices at $\beta=6.0,6.2,6.8$. Plot (b) compares the ``hypercubic-free 
 propagator'' defined from eqs.~(\ref{roies},\ref{pentes}) (black circles) and the propagator 
 computed by direct extrapolation to 
$a^2p^{[4]}=0$ (white circles), plotted as a function of the momentum.}}
%%%%%%%%%%%%%%%%%%%%%%%%%%%%%%%%%%%%%%%%%%%%%%%%%%%%%%%%%%%%%%%
\end{center}
\end{figure}

  
A simple dimensional argument leads to: 
$\left. {\partial G/( a^2 \partial p^{[4]})}
\right|_{a^2 p^{[4]}=0} \propto 1/p^4$, as $L \to \infty$, suggesting the fitting form 
$b/p^4$, where $b$ is a constant \footnote{For example, the hypercubic  
correction for the free propagator, $1/{\widetilde p^2}=1/(p^2 - 
\frac 1 {12} a^2 p^{[4]})+ \dots $, would be ${1\over 12 p^4}$}. 
At this point, from the study of the lattice hypercubic slope 
for different values of $\beta$, we extracted the values of term $b/p^4$ and examined the behaviour 
of the remainder. 
We find that the logarithm of $\left( -b + p^4 \left. {\partial G/( a^2 \partial p^{[4]})} 
\right|_{a^2 p^{[4]}=0} \right)$ is well described by a linear function of $Lp$. This leads 
us to:
\bea
\left. {\partial G\over \partial (a^2 p^{[4]})} \right|_{a^2 p^{[4]}=0} = \ {b\over p^4}\ 
\biggl(1+ c \ 
\exp(-d \ L p) \biggr) \ .
\label{pentes}
\eea 
By fitting our data to the above form and by keeping $\chi^2/d.o.f. \simeq 1$ 
(see Fig.\ref{figpentes}a), 
we obtained the following set of parameters:
\begin{table}[h]
\begin{center}
\begin{tabular}{|c|ccc|}\hline 
{\phantom{\Huge{l}}}\raisebox{-.1cm}{\phantom{\Huge{j}}}
$\beta$ & $b$ & $c$ & $d$\\ \hline
{\phantom{\Huge{l}}}\raisebox{-.1cm}{\phantom{\Huge{j}}}
6.0 &0.117(9)&1.1(1.5)& 0.15(8)  \\
{\phantom{\Huge{l}}}\raisebox{-.1cm}{\phantom{\Huge{j}}}
6.2 &0.109(2)&1.9(1.2)& 0.18(3)  \\
{\phantom{\Huge{l}}}\raisebox{-.1cm}{\phantom{\Huge{j}}}
6.8 &0.100(5)&0.6(8)& 0.13(5)  \\ \hline
\end{tabular}
\end{center}
\caption{\label{parapentes}{\sl 
The values of the parameters $b$, $c$, $d$,
as obtained by fitting our data to eq.\rm{(\ref{pentes}}).}}
\end{table}

\noindent
The values for $c$ and $d$ in Tab.~\ref{parapentes} are 
presented to show the order of magnitude. Their errors, 
estimated by using the jackknife method, are misleading since the parameters are strongly correlated. 
A refined statistical study is not necessary for our purpose, 
since we follow the jackknife analysis cluster by cluster to the end.

%%For our final fit, we will use only the points where the finite volume corrections to
%%the slope ( the $\exp(-d \ L p)$ term )  are small, {\it i.e.} $\mu > 2.8$~GeV 
%%for $\beta=6.0$, $6.2$, and $\mu>5.6$~GeV, for $\beta=6.8$. In spite of their 
%%smallness they play an 
%%important role to insure, for example, the scaling in $a$ and the 
%%smoothness of the final results.  

On the other hand it is rewarding that the errors on $b$ in (\ref{parapentes}) are small, which is 
essential for an accurate infinite volume limit. It is also encouraging that $b$ varies only slowly 
with $a$, which justifies our neglecting higher order terms, ${\cal O}(a^4\ p^{[6]})$ etc., and it clearly 
confirms that we do control the lattice hypercubic artifacts. 
We can now take advantage of the good fit obtained with eq.~(\ref{pentes}) and compute the {\sl 
lattice hypercubic slope} in cases where only one orbit exists (for instance, when $n^2=5$, we only 
have $n_{\mu}=(2,1,0,0)$ and its $H_4$ orbit).

%%%%%%%%%%%%%%%%%%%%%%%%%%%%%%%%%%%%%%%%%%%%%%%%%%%%%%%%%%%
%%%%%%%%%%%%%%%%%%%%%%%%%%%%%%%%%%%%%%%%%%%%%%%%%%%%%%%%%%%
%%%%%%%%%%%%%%%%%%%%%%%%%%%%%%%%%%%%%%%%%%%%%%%%%%%%%%%%%%%
%%%%%%%%%%%%%%%%%%%%%%%%%%%%%%%%%%%%%%%%%%%%%%%%%%%%%%%%%%%

To summarize, by using eq.~(\ref{pentes}), we extrapolate $G_{\rm lat}^{(2)}(p^2,a^2 p^{[4]};L,a)$ 
to what we will call the {\sl hypercubic-free propagator}, $G_{\rm lat}^{(2)}(p^2,0;L,a)$. 
In ref.~\cite{propag}, we discussed the improvement brought in by our previous 
($a^2p^{[4]}=0$)-extrapolation 
approach with respect to other methods to reduce hypercubic artifacts. 
The use of eq.~(\ref{pentes}) allows even better accuracy on slopes: not only does it allow an 
extrapolation to $a^2p^{[4]}=0$ when only one orbit exists but it also helps 
to reduce the uncertainty by taking benefit of the neighboring values of $n^2$ when the error is 
locally too
large. The outcome of this improvement is depicted in Fig.~\ref{figpentes}b, where the resulting 
curve joining the points obtained by using eq.~(\ref{pentes}) is much smoother than the one obtained 
by separate extrapolation~\footnote{Note however, that even in the latter case, the resulting curve 
is by far smoother than the one obtained by simply averaging the orbits or by selecting the 
``democratic'' points, as advocated in ref.~\cite{leinweber}. This improvement was already illustrated 
in ref.~\cite{propag}.} for each $n^2$.

It is important to add that not all ${\cal O}(a^2)$ artifacts are eliminated: 
for instance, the 
lattice artifacts $\propto a^2 \ p^2$, which do not break $SO(4)$ 
invariance, are still present. 
One way to deal with this problem is to compare the 
data at different values of $\beta$.
One can also study the stability of the results when the maximum
value of $p^2$ used in the fits is varied. 
We have checked this stability in the fits presented below.

\subsection{Finite volume effects}

\begin{figure}[t!]
%\vspace*{-1.cm}
\begin{center}
\begin{tabular}{@{\hspace{-1.0cm}}c c c}
(a) &\hspace*{-0.5cm} & (b) \\
\epsfxsize8.8cm\epsffile{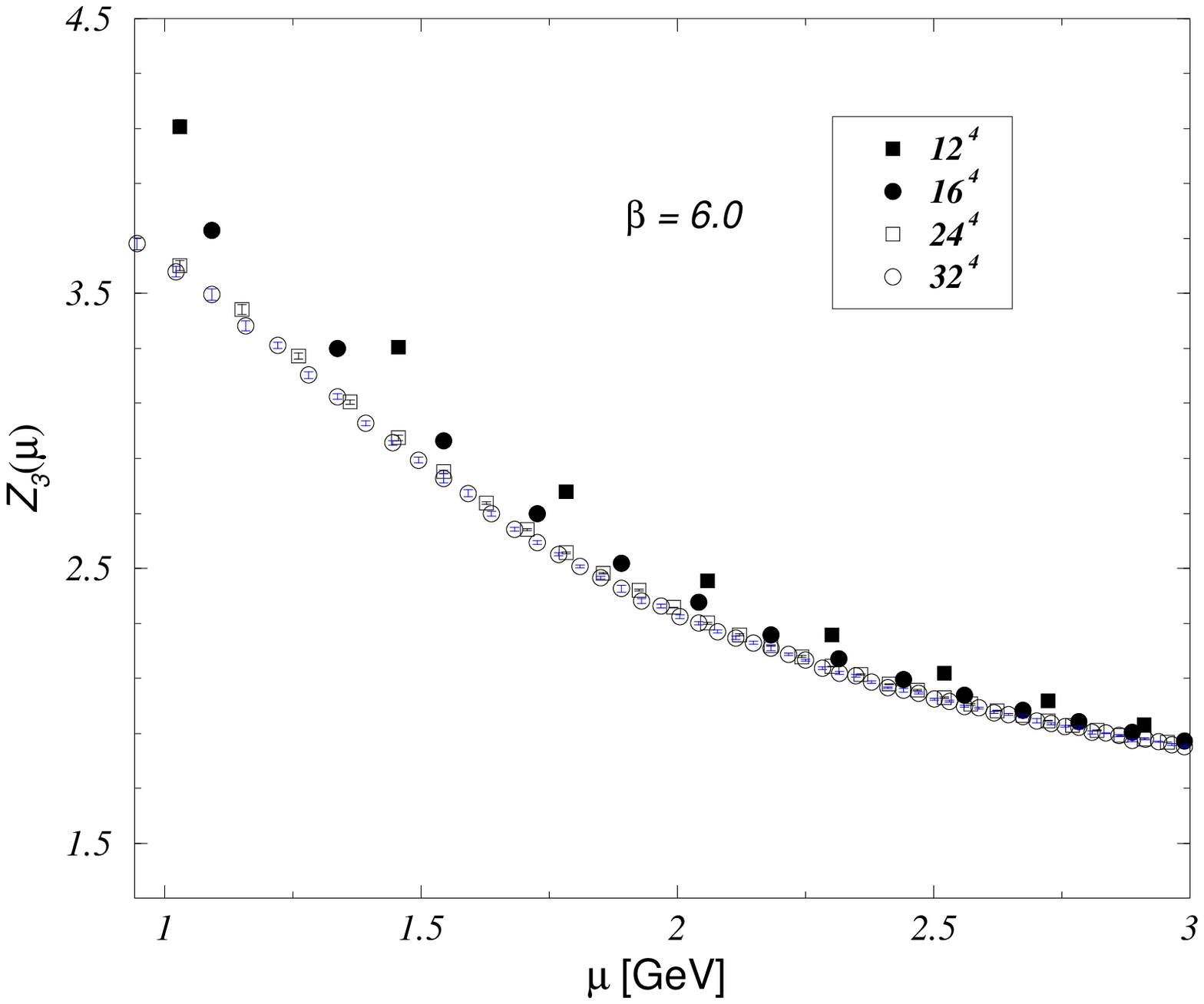} &\hspace*{-0.5cm} & \epsfxsize8.8cm\epsffile{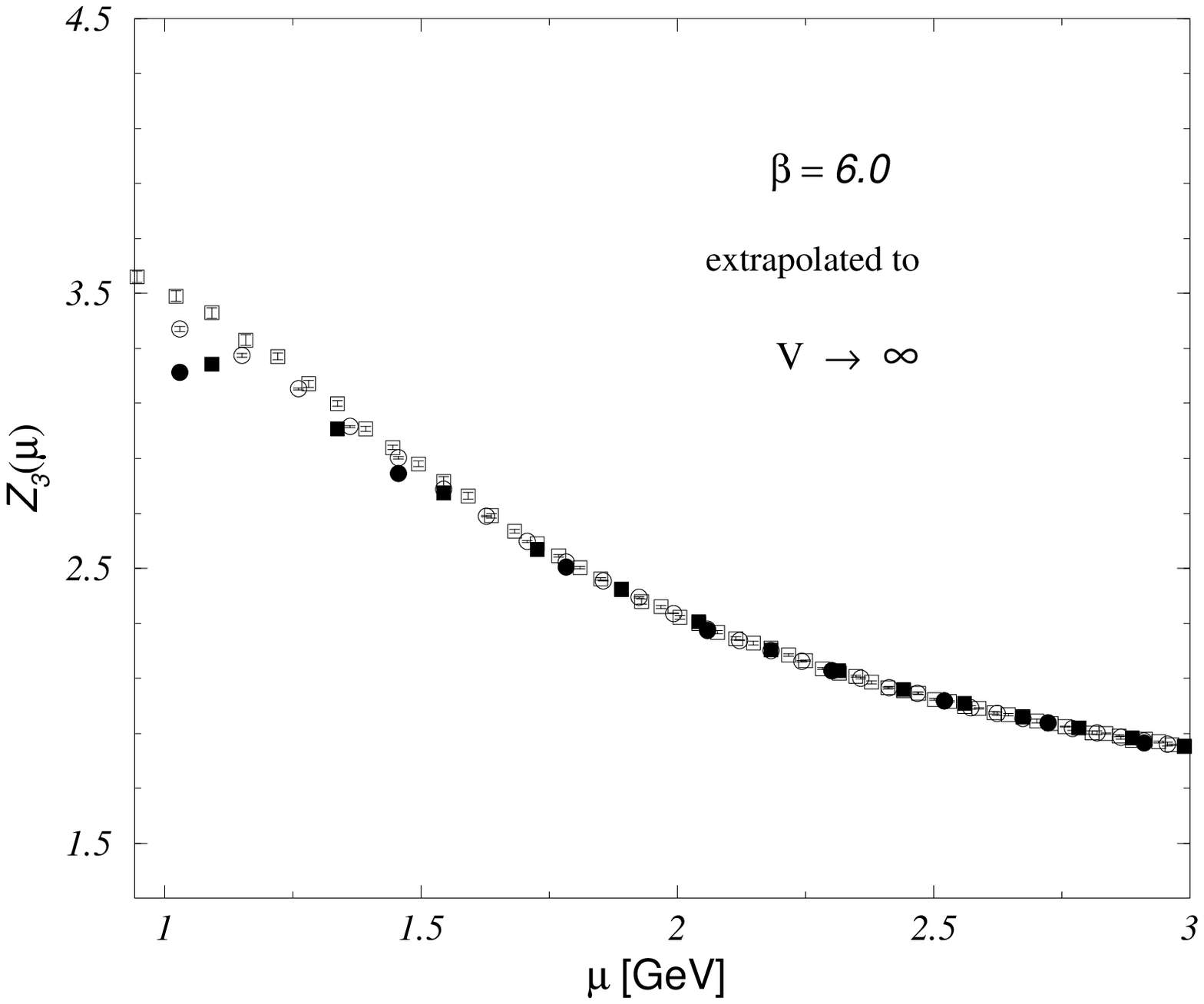}  \\
\end{tabular}
%%%%%%%%%%%%%%%%%%%%%%%%%%%%%%%%%%%%%%%%%%%%%%%%%%%%%%%%%%%%%%%%%%
\caption{\label{figvol}{\sl 
Plot (a) contains the {\it hypercubic-free propagators} evaluated on $12^4$ (black
squares), $16^4$ (black circles), $24^4$ (white circles), $32^4$ (white squares) lattices at 
$\beta=6.0$. The same data extrapolated to $L \to \infty$ according to 
the parametrization given 
by eqs.~(\ref{vol},\ref{volparameters}) are plotted in (b)  
.}}
%%%%%%%%%%%%%%%%%%%%%%%%%%%%%%%%%%%%%%%%%%%%%%%%%%%%%%%%%%%%%%%%%%
\end{center}
\end{figure}

After removing hypercubic artifacts, we are left with the {\sl hypercubic-free propagator}. The 
dependence on the length scale, $L$, as we previously mentioned, is apparent from Fig.~\ref{figvol}. 
The elimination of this additional length scale should be done in order to compute the renormalization 
constant for the Landau gauge gluon propagator,
\bea
Z_3(\mu,1/a) \ = \ \lim_{L \to \infty} 
\bigl( p^2  G_{\rm lat}^{(2)}(p^2,0;L,a)
\bigr)_{p^2=\mu^2} \ .
\label{zmom}
\eea

\noindent  In doing so, we will not attempt a theoretical understanding of the 
expected finite volume dependence of the bare propagator. We will be content if we obtain a reliable
 empirical 
parametrization for the dependence on the lattice volume of $G_{\rm lat}^{(2)}(p^2,0;L,a)$ which will 
allow us to take the required limit~(\ref{zmom}). For dimensional reasons, we will take it as a function 
of $L p$ and $a/L$. We note that the difference amongst the data at fixed $\beta$ and various volumes 
gets bigger as we move towards lower $p^2$. This is illustrated in Fig.~\ref{figvol} where we plot 
$p^2 G_{\rm lat}^{(2)}(p^2,0;L,a)$. We tried several {\it ans\"atze} for this parametrization function 
and finally found that

\beq
G_{\rm lat}^{(2)}(p^2,0;L,a)\ =\ G_{\rm lat}^{(2)}(p^2,0;\infty,a)\ \left( 1 + v_1 \left({a\over
L}\right)^4 + v_2 \exp(-v_3 L p)\right) \ ,
\label{vol}
\eeq

\noindent with 

\beq
v_1=450(40), \;\;\;\; v_2=0.44(16), \;\;\; v_3=0.177(25) ,
\label{volparameters}
\eeq

\noindent gives the best fit to the behaviour on $L$ of {\sl hypercubic-free propagators} evaluated 
on $12^4$, $16^4$, $24^4$, $32^4$ lattices in the energy window~\footnote{In this procedure we assume that the volume $32^4$ is already infinite, {\it i.e.} we take $G_{\rm lat}^{(2)}(p^2,0;32,a(6.0)) 
\simeq G_{\rm lat}^{(2)}(p^2,0;\infty,a(6.0))$. In the fit to the form~(\ref{vol}), the energy window is chosen such that the total $\chi^2/d.o.f. \sim 2$.} ($1.5\div 3.0)$~GeV at $\beta=6.0$. 
This parametrization is not efficient at lower energies, 
as it can be seen in Fig.~\ref{figvol}b. 

%%%%%%%%%%%%%%%%%%%%%%%%%%%%%%%%%%%%%%%%%%%%%%%%%%%%%%%%%%%
%%%%%%%%%%%%%%%%%%%%%%%%%%%%%%%%%%%%%%%%%%%%%%%%%%%%%%%%%%%
%%%%%%%%%%%%%%%%%%%%%%%%%%%%%%%%%%%%%%%%%%%%%%%%%%%%%%%%%%%

\begin{figure}[t!]
%\vspace*{-1.cm}
\begin{center}
\begin{tabular}{@{\hspace{-1.3cm}}c c c}
(a)& &(b) \\
\epsfxsize8.8cm\epsffile{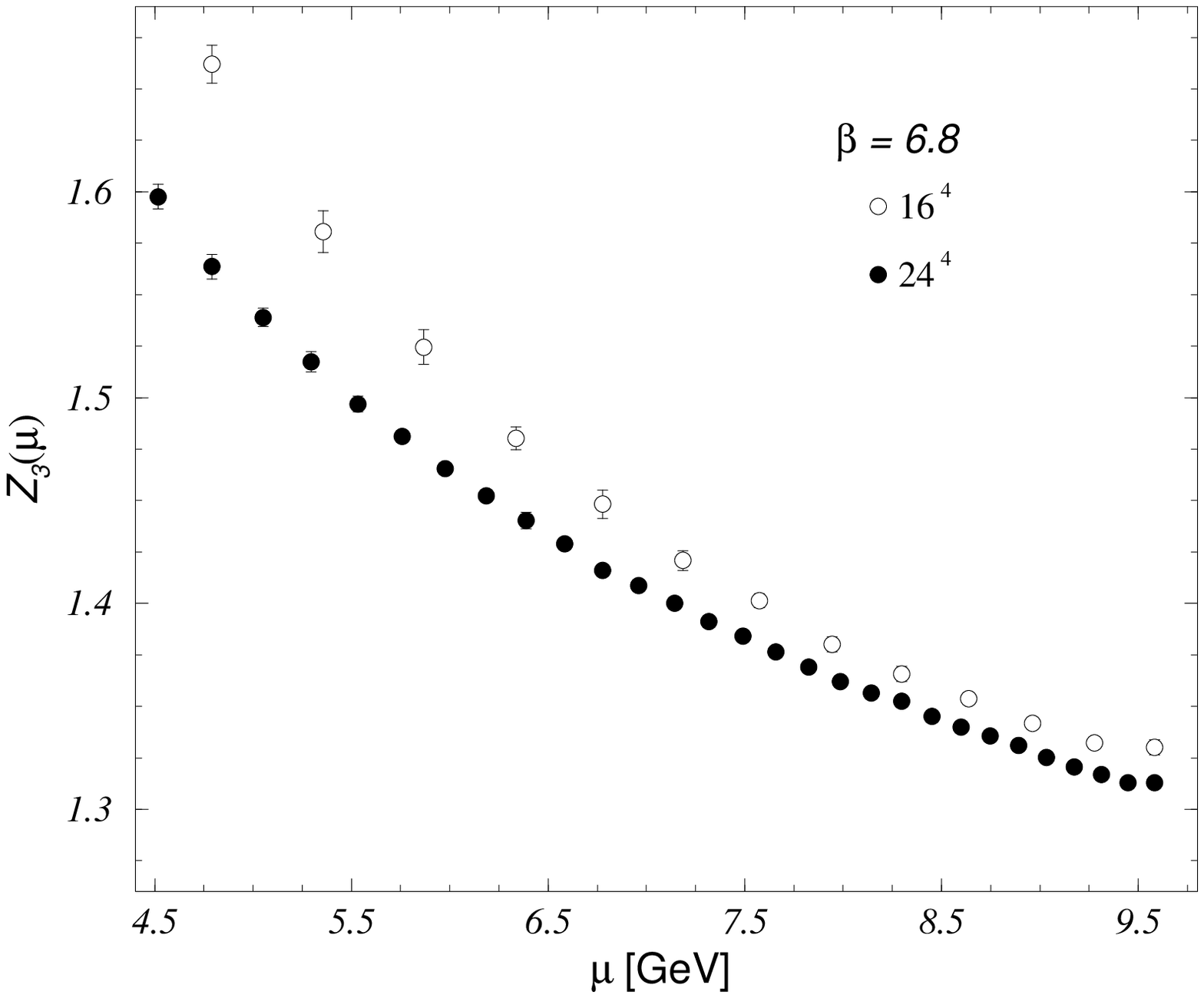} &\hspace*{-0.5cm} & \epsfxsize8.8cm\epsffile{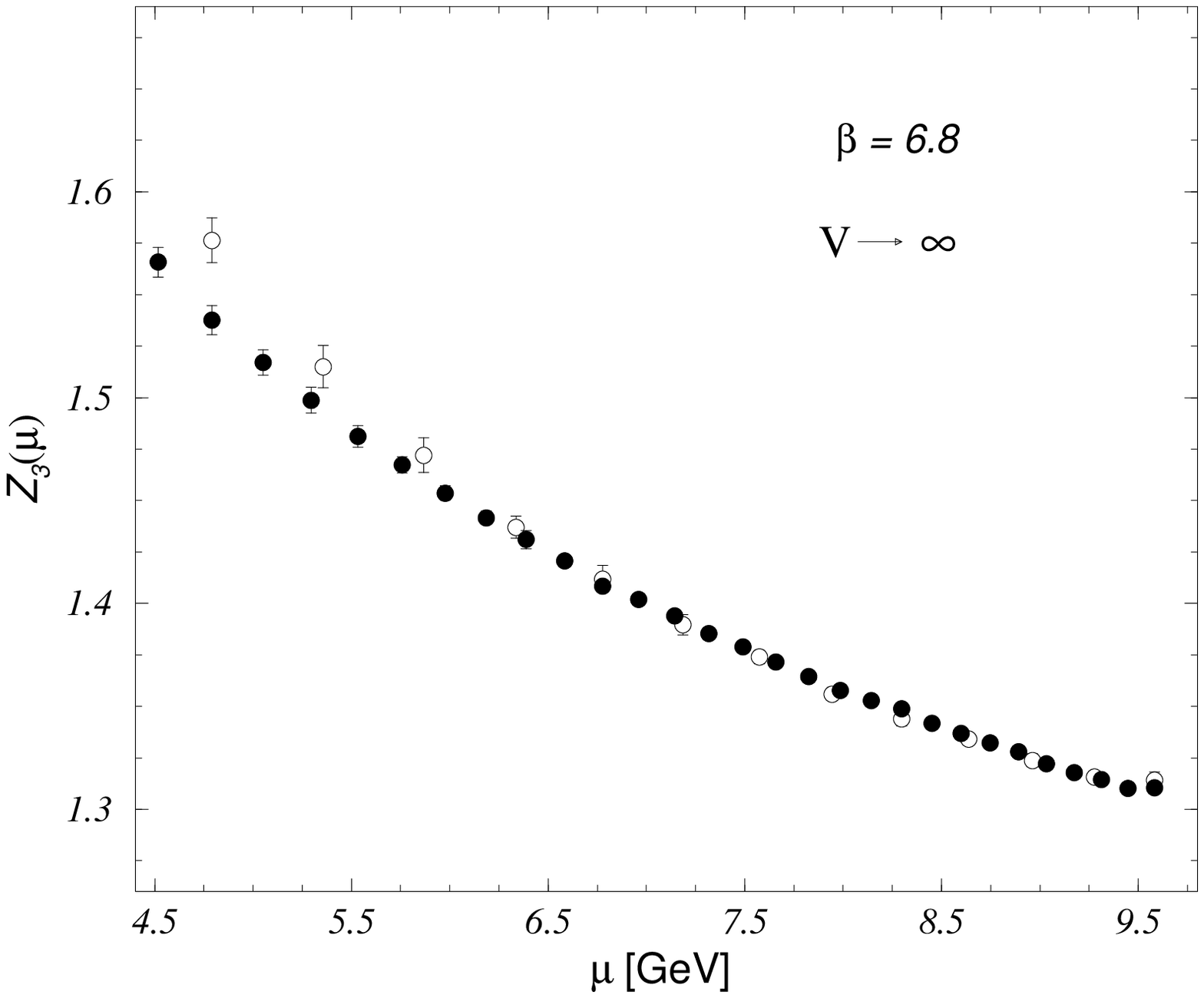}  \\
\end{tabular}
%%%%%%%%%%%%%%%%%%%%%%%%%%%%%%%%%%%%%%%%%%%%%%%%%%%%%%%%%%%%%%%%%%

\caption{\label{figvol2}{\sl 
This figure is analogous to Fig. \ref{figvol} for $\beta=6.8$. Black (white)
circles correspond to $24^4$ ($16^4$) lattices.}}
%\label{figpentes}
%%%%%%%%%%%%%%%%%%%%%%%%%%%%%%%%%%%%%%%%%%%%%%%%%%%%%%%%%%%%%%%%%%
\end{center}
\end{figure}

Once the parametrization function from propagators evaluated at $\beta=6.0$ is established, it can 
be applied to our results at $16^4$ and $24^4$ lattices at $\beta=6.8$. The
agreement after extrapolation shown by the curves resulting from the extrapolation in 
Fig.~\ref{figvol2} is a crucial test
for the validity of such a parametrization for the finite volume effects. 

%%%%%%%%%%%%%%%%%%%%%%%%%%%%%%%%%%%%%%%%%%%%%%%%%%%%%%%%%%%
%%%%%%%%%%%%%%%%%%%%%%%%%%%%%%%%%%%%%%%%%%%%%%%%%%%%%%%%%%%
%%%%%%%%%%%%%%%%%%%%%%%%%%%%%%%%%%%%%%%%%%%%%%%%%%%%%%%%%%%

Thus, eqs.~(\ref{zmom}-\ref{volparameters}) lead to a non-perturbative evaluation of the renormalization constant, $Z_3(\mu,1/a)$, which is precisely the quantity we want to compare to the predictions of perturbative QCD.

\section{Fitting in the  $\widetilde {\rm MOM}$ scheme at high energies}

In this section we perform the matching of the perturbative predictions for $\widetilde {\rm MOM}$ 
renormalization constant to our lattice result at high energies. We follow the method outlined in 
section~\ref{method} and refer the reader to ref.~\cite{propag} for more details.

We consider the coupled differential equations~(\ref{gamma},\ref{beta}) in the $\widetilde {\rm MOM}$ 
scheme, where the coefficients $\gamma_1$, $\gamma_2$, $\beta_2$ are those given in eq.~(\ref{beta2}). 
We fit our data at $\beta=6.8$ for $Z_3(\mu,1/a)$, with a solution of these coupled equations in the 
energy window $( 5.6 \div 9.5 )$~GeV. The result of this fit is
\beq
Z_3(9.5\ {\rm GeV})=1.3107(9),\quad \alpha_{\widetilde{\rm MOM}}(9.5\ {\rm GeV})=0.190(3)
 ,\quad \chi^2/{ d.o.f.} = 0.29 \,. \label{resmom}
\eeq

\noindent The $\chi^2/d.o.f$ is significantly smaller than $1$ which may be a sign of some correlation 
between the points at different values of the energy $\mu$. 

\begin{figure}[t!]
%\vspace*{-1.cm}
\begin{center}
\hspace*{.8cm}~\epsfxsize10.5cm\epsffile{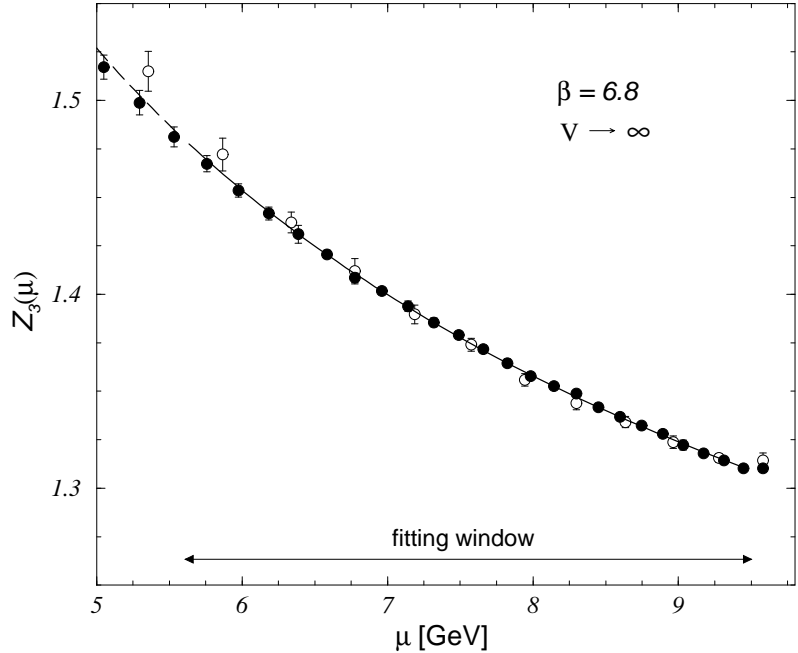}
%%%%%%%%%%%%%%%%%%%%%%%%%%%%%%%%%%%%%%%%%%%%%%%%%%%%%%%%%%%%%%%%%%

\caption{\label{figfit1}{\sl The plot shows (full line) the best fit to $Z_3(\mu,1/a)$
with the three-loop formula  together with
the lattice results at $\beta=6.8$ after extrapolation to $L \to \infty$ from both 
$16^4$ and $24^4$ lattice volumes, for $5.6 \ {\rm GeV} < \mu < 9.5 {\rm GeV}$. The fit
is continued outside the energy window as a dashed line.}}

%%%%%%%%%%%%%%%%%%%%%%%%%%%%%%%%%%%%%%%%%%%%%%%%%%%%%%%%%%%%%%%%%%
\end{center}
\end{figure}

As explained in~\cite{propag}, $\Lambda^{\overline {\rm MS}}$ can be estimated from the above 
quantities~(\ref{resmom}), by using the perturbative expressions to two- and three-loop accuracy. We 
obtain:
\bea
&&\Lambda_{(3 \rm{loop})}^{\overline {\rm MS}}\ \simeq\ 0.346\; 
\Lambda_{(3 \rm{loop})}^{\widetilde {\rm MOM}}\ =\ 319 \pm 14 \  {\rm MeV},\quad\cr
&&\hfill \cr
&&\Lambda_{(2 \rm{loop})}^{\overline {\rm MS}}\ \simeq\ 0.346\; 
\Lambda_{(2 \rm{loop})}^{\widetilde {\rm MOM}}\ \simeq\ 375  \ {\rm MeV},
\label{lambdamom}
\eea
where the error is only statistical at this stage. The existence of a good fit 
(see Fig. \ref{figfit1}) 
by itself is not a sufficient proof of asymptoticity: next-to-three-loop corrections 
could be mimicked 
by a simple rescaling of $\Lambda_{\rm QCD}$ in the considered energy range~\cite{propag}. 
This is why we developed a consistent method to test asymptoticity by exploring the 
scheme dependence within the domain of so-called ``{\it good schemes}'':
 one investigates the dispersion of the result for 
$\Lambda^{\overline{\rm MS}}$  
when we vary the schemes , by varying $\gamma_1$ and $\gamma_2$,  
 in all the possible ways such  that 
the successive terms in the perturbative series~(\ref{gamma},\ref{beta}) 
are at most as large as the preceding ones~\footnote{This is the generalization of 
the effective charge approach proposed in ref.~\cite{grunberg}.}.  
In ref.~\cite{propag}, we found that this dispersion is of $\sim 35$~MeV, 
for $\Lambda^{\overline{\rm MS}}$ fixed from the fit of 
the gluon propagator at $\sim 4$~GeV to the three-loop perturbative expression. 
In the present study, when fixing $\Lambda^{\overline {\rm MS}}$ at around $9.5$~GeV, 
this dispersion (in the same domain of schemes) is of $\sim 10$~MeV only.

Note that the difference in eq.~(\ref{lambdamom}) between two and three loops, although smaller than at $4$~GeV, is still sizable indicating the necessity to include the third loop term. 

%%%%%%%%%%%%%%%%%%%%%%%%%%%%%%%%%%%%%%%%%%%%%%%%%%%%%%%%%%%
%%%%%%%%%%%%%%%%%%%%%%%%%%%%%%%%%%%%%%%%%%%%%%%%%%%%%%%%%%%
%%%%%%%%%%%%%%%%%%%%%%%%%%%%%%%%%%%%%%%%%%%%%%%%%%%%%%%%%%%

\section{Global description and asymptotic pattern}
\label{pattern}

Our analysis at high energy in the previous section seems to establish that a signal 
of three-loop 
perturbative scaling is found in the energy window $(5.6 \div 9.5)$~GeV. By including 
the data obtained at $\beta=6.0$ and $\beta=6.2$, one can make the energy window larger: 
$2.8$ GeV$\lesssim 
\mu \lesssim 9.5$ GeV (the choice for the lower limit will be discussed below). 
However, as already mentioned in the end of Sec.~\ref{method}, any global fit
involves two additional parameters,
the ratios $Z_3^b(6.0)/Z_3^b(6.8)$ and $Z_3^b(6.2)/Z_3^b(6.8)$.
Moreover, we do not expect a three-loop perturbative behaviour to work
in the whole energy window. In ref.~\cite{propag}, we showed that the
propagator was not asymptotic to three-loops at $\sim 4$~GeV. 
The difference between $\Lambda_{(3\rm{loop})}^{\overline
{\rm MS}}$ in eq.~(\ref{lambdamom}) and $\sim 350$~MeV as found in
ref.~\cite{propag}, confirms that statement. 
Thus, at least the fourth loop correction is necessary for  the global fit. 
Unfortunately, such a perturbative result is not available in the
${\widetilde {\rm MOM}}$ scheme.

With these fives free parameters, a global fit turns out to be 
unstable. A global study of our lattice data would nevertheless enable 
 a direct test of consistency for the the whole information we extract
 from gluon propagator. For that reason we adopt the following strategy.
First, we take the value of $\Lambda^{\overline {\rm MS}}_{\rm (3loop)}$ 
given in eq.~(\ref{lambdamom}), to be the asymptotic one, i.e. we assume
the gluon propagator to reach asymptotia at three-loop for the energy
 window studied in the previous section. Then we fix the 
fourth loop correction to the
three-loop perturbative expression, by fitting the data obtained in our 
simulations at $\beta=6.0$ and $\beta=6.2$, corresponding
to the energy range $(2.8 \div 4.3)$~GeV. Once, the fourth loop 
correction is known, we will
verify the asymptoticity of the gluon propagator in the entire energy 
window ($2.8 \div 9.5$)~GeV.

The four-loop information about the gluon propagator in ${\widetilde{\rm MOM}}$-scheme is 
encoded in the coefficients $\gamma_3^{\widetilde{\rm MOM}}$ and $\beta_3^{\widetilde{\rm MOM}}$. 
These two coefficients are not independent but related through 
the expression~\cite{beta2,propag}
\beq
\frac{\gamma_3^{\widetilde{\rm{MOM}}}}{(4\pi)^4} + \frac{\beta_3^{\widetilde{\rm{MOM}}}}{(4\pi)^4} 
\frac{\gamma_0}{\beta_0}=\frac{\gamma_3}{(4\pi)^4} + \frac{\beta_3}{(4\pi)^4} 
\frac{\gamma_0}{\beta_0} \ ,
\label{condition}
\eeq 
\noindent which is valid for any renormalization scheme in which 
$\gamma_1=\gamma_1^{\widetilde{\rm{MOM}}}$, $\gamma_2=\gamma_2^{\widetilde{\rm{MOM}}}$
 and $\beta_2=\beta_2^{\widetilde{\rm{MOM}}}$, listed in eq.~(\ref{beta2}).
Thus, there is only one free parameter to be fitted. For simplicity, 
we choose among the set of schemes satisfying ~(\ref{condition}), 
the one with $\beta_3=0$, $\widehat{\gamma_3}$ being that free parameter. For such a renormalization scheme,
\beq
Z^R(\mu)\ =\ Z^{(3)}(\mu)\ \exp\left(\frac{1}{3(4\pi)^3}\frac{\widehat{\gamma}_3}{\beta_0}
\ (\widehat{\alpha}^3(\mu)-
\alpha^3_0 \ ) \ 
\right)
\label{fourloop}
\eeq

%%%%%%%%%%%%%%%%%%%%%%%%%%%%%%%%%%%%%%%%%%%%%%%%%%%%%%%%%%%
%%%%%%%%%%%%%%%%%%%%%%%%%%%%%%%%%%%%%%%%%%%%%%%%%%%%%%%%%%%
\begin{figure}
%\vspace*{-1.cm}
\begin{center}
\begin{tabular}{@{\hspace{-1.0cm}}c c c}
(a) & \hfill   & \hfill \\  
\hfill & \epsfxsize10.0cm\epsffile{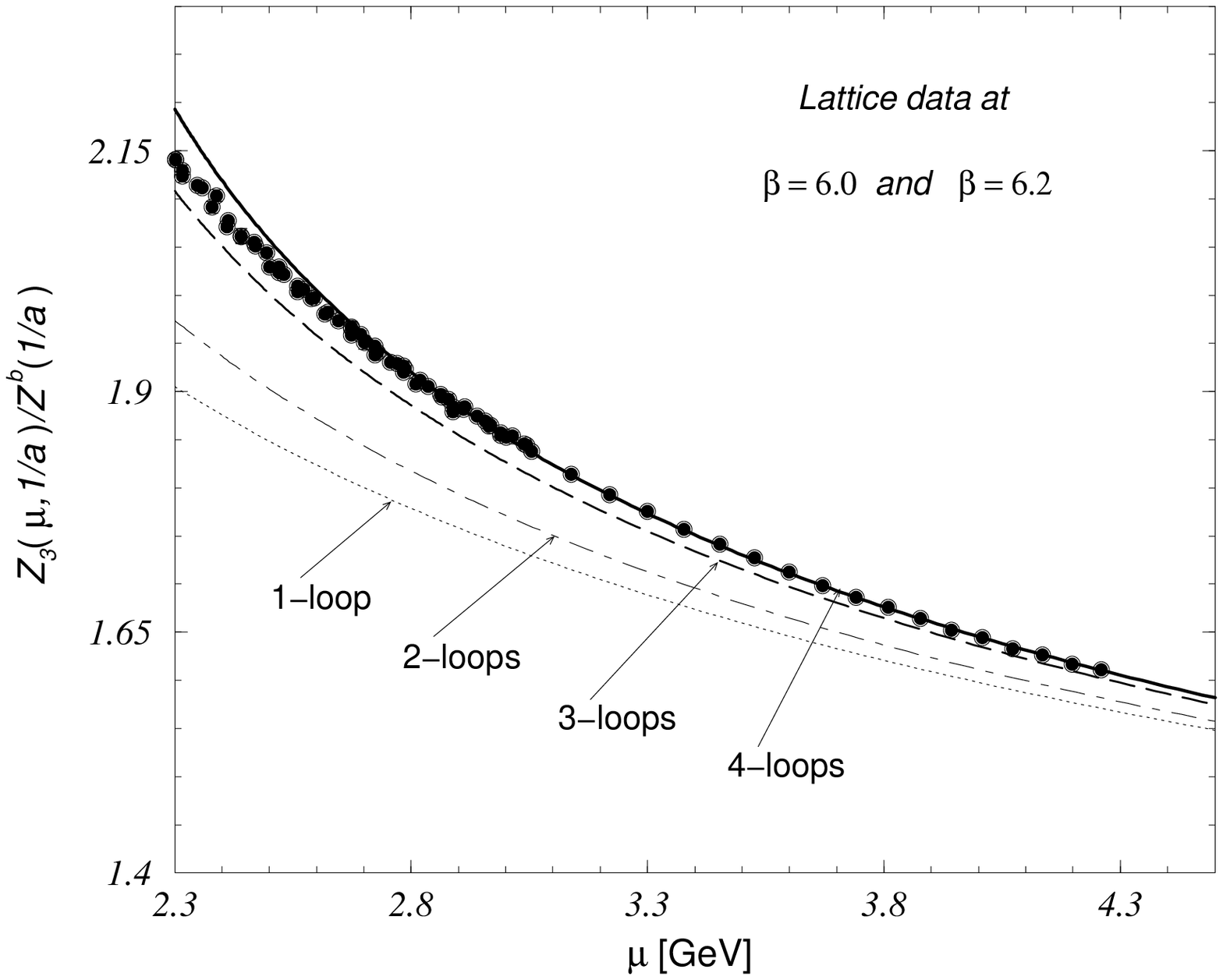}  & \hfill \\  
(b) & \hfill   & \hfill \\  
\hfill & \epsfxsize10.0cm\epsffile{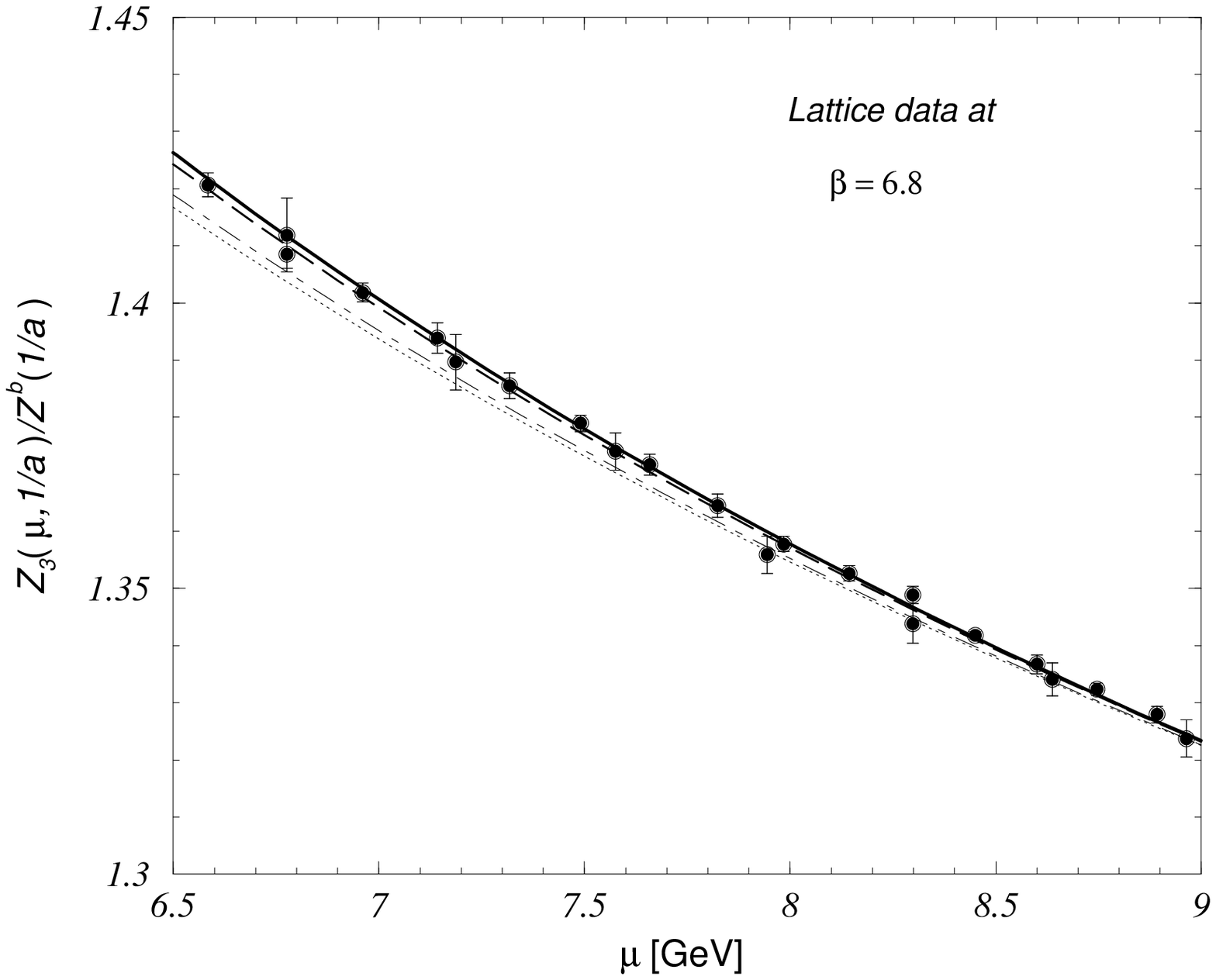} &\hfill 
\end{tabular}
%%%%%%%%%%%%%%%%%%%%%%%%%%%%%%%%%%%%%%%%%%%%%%%%%%%%%%%%%%%%%%%%%%

\caption{\label{total1}{\sl One-, two-, three- and four-loop perturbative curves obtained
 from our best fits of $\alpha_{\widetilde{\rm{MOM}}}(9.5 \rm{GeV})$ and $\widehat{\gamma}_3$ are
presented in plot (a).
 Notice that all these curves are computed with 
 $\Lambda^{\widetilde {\rm MOM}}\ =\ 319 \pm 14 \  {\rm MeV}$.
  The points correspond to the lattice evaluations at $\beta=6.0$, $6.2$ 
  divided by our best fits 
of the ratios of $Z^b_3$ referred to $\beta=6.8$ ($Z^b_3$ 
is taken to be $1$ at $\beta=6.8$). 
Plot (b)
 shows the same
perturbative curves and the lattice data at $\beta=6.8$. }}
%%%%%%%%%%%%%%%%%%%%%%%%%%%%%%%%%%%%%%%%%%%%%%%%%%%%%%%%%%%%%%%%%%
\end{center}
\end{figure}
\begin{figure}
%\vspace*{-1.cm}
\begin{center}
\begin{tabular}{@{\hspace{-1.0cm}}c c c}
(a)& & \\
 &\epsfxsize9.8cm\epsffile{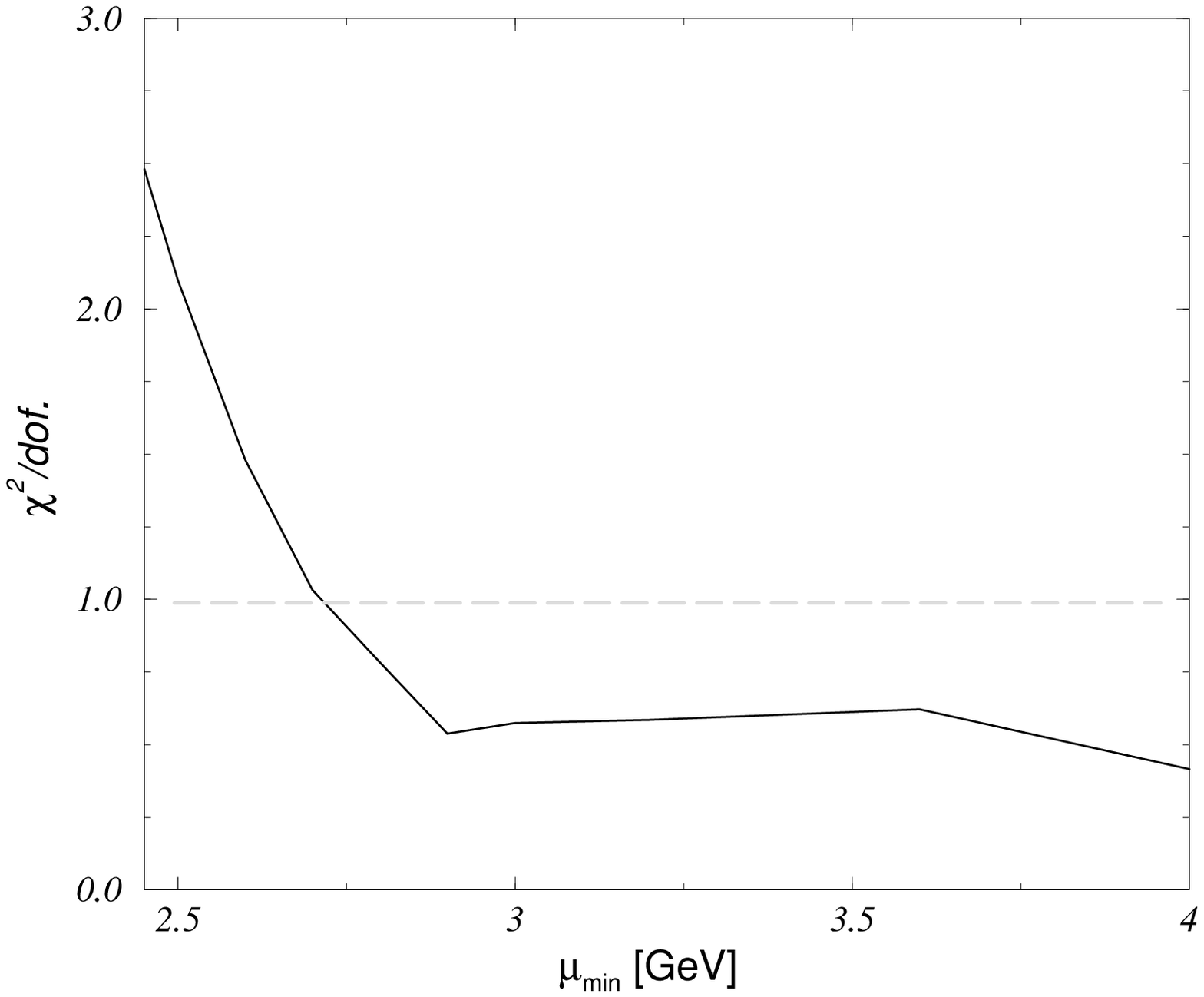} & \\
(b)& & \\
 &\epsfxsize9.8cm\epsffile{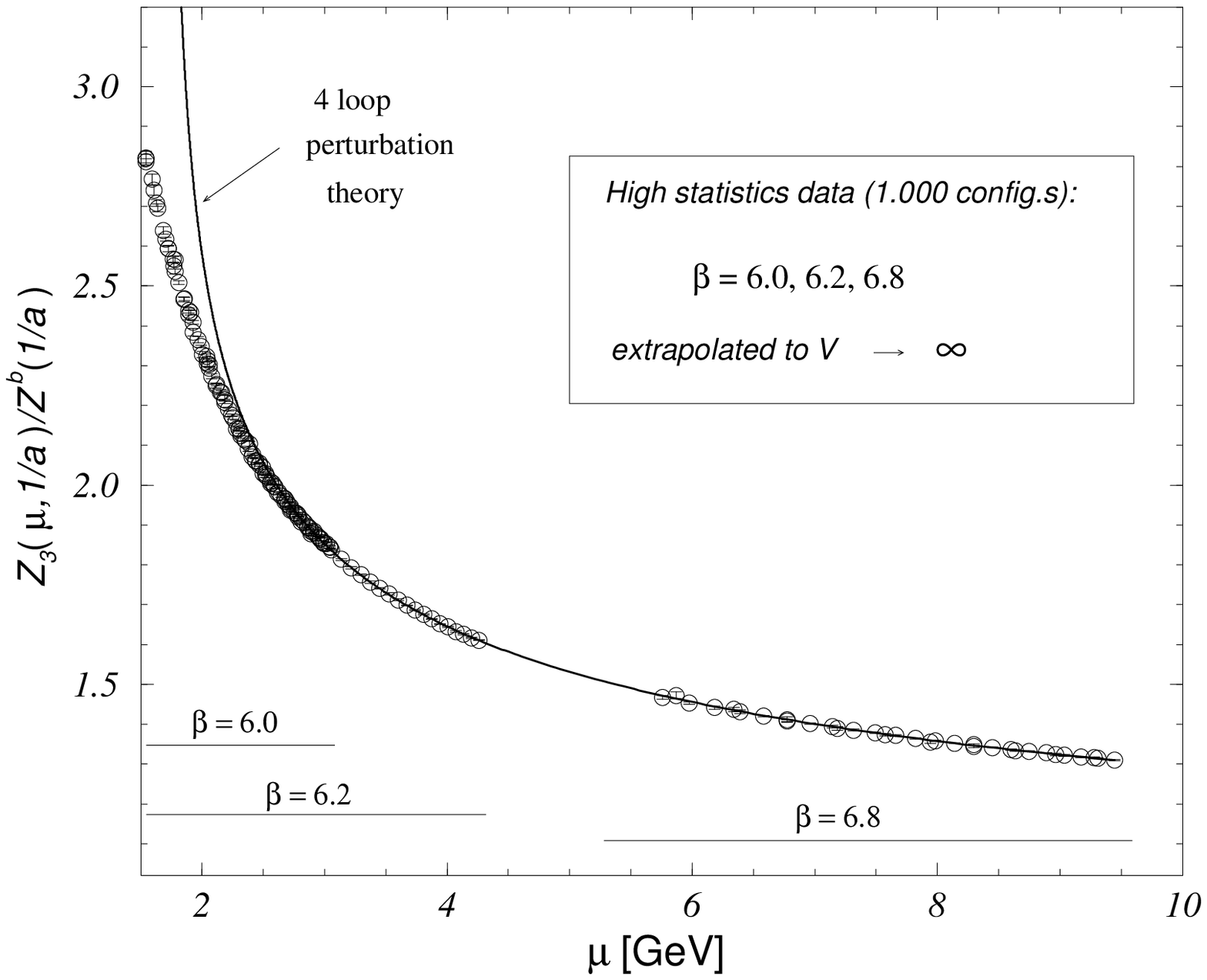} &  \\
\end{tabular}
%%%%%%%%%%%%%%%%%%%%%%%%%%%%%%%%%%%%%%%%%%%%%%%%%%%%%%%%%%%%%%%%%%

\caption{\label{total2}{\sl Plot(a): the $\chi^2/{\rm d.o.f.}$ for the global fit obtained after the 
estimation of the four-loop contribution as a function of the lower limit for the energy window. 
Plot (b): the global
fit of all the lattice results with the four-loop corrected perturbative formula (the full line curve on
Fig.~\ref{total1}) from eq.~(\ref{fourloop}).}}
%%%%%%%%%%%%%%%%%%%%%%%%%%%%%%%%%%%%%%%%%%%%%%%%%%%%%%%%%%%%%%%%%%
\end{center}
\end{figure}

%%%%%%%%%%%%%%%%%%%%%%%%%%%%%%%%%%%%%%%%%%%%%%%%%%%%%%%%%%%
\noindent is, up to higher irrelevant orders, the solution to the four-loop coupled equations analogous 
to~(\ref{gamma}) and (\ref{beta}),
where $Z^{(3)}(\mu)$ is the solution of the three-loop problem, with $\alpha_0$ 
being the initial
strong coupling constant at $\mu_0$ for both, three- and four-loops. 
 The results of the fit in the energy window $(2.8 \div 4.3)$~GeV read 
(see Fig.~\ref{total1})

\bea
\frac{Z^b_3(a(6.0))}{Z^b_3(a(6.8))}\ =\ 0.995(3)\ ,\quad \; \frac{Z^b_3(a(6.2))}{Z^b_3(a(6.8))}
\ =\ 1.012(2)\ , \nonumber \\
\hfill \nonumber \\
  \widehat{\gamma}_3\ =\  (2.2 \pm 1.6)  \cdot 10^4
   \quad \quad \left(\chi^2/{d.o.f.} = 1.17\right)\,;
\label{resZb}
\eea

\noindent where the errors on the ratios of $Z^b_3$'s do not take into account the uncertainty coming
from the errors on the lattice spacing ratios. This uncertainty  does not  exceed one percent.
When the lower limit of the energy window takes values below 
$2.8$~GeV, the $\chi^2/d.o.f.$ 
rapidly increases, which indicates the end of the four-loop matching. 
This is illustrated in Fig.~\ref{total2}a. The upper limit is fixed by $a p \le \pi/2$. 
 We note also that the fitted values of the ratios~(\ref{resZb}) are very close to $1.0$, 
 and somewhat larger than the results obtained by using the one-loop lattice perturbation 
 theory~\footnote{For comparison, the one-loop perturbative values of the 
 two ratios are : $Z^b_3(6.0)/Z^b_3(6.8) =0.929$, and  $Z^b_3(6.2)/Z^b_3(6.8) = 0.947$. 
 in clear disagreement with the fitted values (eq. (\ref{resZb})
 even if the small uncertainty on the lattice spacing is considered. On the contrary,
 the fitted ratio $Z^b_3(6.2)/Z^b_3(6.0) = 1.017$  is in 
 good agreement with the one loop perturbative prediction, 1.019.}.   

The estimated $\widehat{\gamma}_3$, is not as large as it might look: 
\bea
{\widehat{\gamma}_3 \ \alpha/ 4\pi\over \gamma_2} \simeq 0.6\,,\quad {\rm and}\quad 
 {\widehat{\gamma}_3 \ \alpha^3/(4\pi)^3\over \gamma_0} \sim 0.05\,
\eea 
at $\mu=4.0$~GeV. The large error for $\widehat{\gamma_3}$ quoted in
 eq.~(\ref{resZb}) is not surprising because the determination of 
 $\widehat{\gamma}_3$ strongly depends on 
 {$\alpha_{\widetilde{\rm MOM}}(9.5\ {\rm GeV})$}. 
 In other words, a very small uncertainty on 
 {$\alpha_{\widetilde{\rm MOM}}$} reflects in large error for
  $\widehat{\gamma}_3$, in our energy window $(2.8 \div 4.3)$~GeV. 
  Thus the computation of the value for $\widehat{\gamma_3}$ from perturbation theory
   would considerably reduce the systematic uncertainty of
    $\alpha_{\widetilde{\rm MOM}}(9.5\ {\rm GeV})$, and hence of 
    $\Lambda^{\overline {\rm MS}}$.

%%%%%%%%%%%%%%%%%%%%%%%%%%%%%%%%%%%%%%%%%%%%%%%%%%%%%%%%%%%
%%%%%%%%%%%%%%%%%%%%%%%%%%%%%%%%%%%%%%%%%%%%%%%%%%%%%%%%%%%
%%%%%%%%%%%%%%%%%%%%%%%%%%%%%%%%%%%%%%%%%%%%%%%%%%%%%%%%%%%

At this point, we can check the consistency by performing a 
global fit over the entire window $2.8\ {\rm GeV} \leq \mu \leq  9.5$~GeV, with the values of 
$\widehat{\gamma}_3$ and of the 
ratios $Z^b_3$ taken from~(\ref{resZb}). The result of such a fit is depicted in Fig.~\ref{total2}. 
The global estimate of $\Lambda^{\overline {\rm MS}}$ does not modify the one we obtained 
at large energies ($5.6\  {\rm GeV} \leq \mu \leq  9.5$~GeV) given in eq.~(\ref{lambdamom}), whereas 
the global $\chi^2/{d.o.f.}$ is $0.79$.

Finally, we should assess the systematic uncertainty introduced by the assumption of the three-loop asymptoticity ({\it i.e.} $\widehat{\gamma_3}=0$), for $5.6\ \rm{GeV}\ \leq \ \mu \ \leq \ 9.5\ \rm{GeV}$. The simplest estimate of the
non-asymptoticity errors consists in monitoring the value of $\Lambda^{\overline {\rm MS}}$, while varying the coefficient $\widehat{\gamma_3} \neq 0$. 
$|\widehat{\gamma_3}|$ is reasonably bounded by $4 \pi 
\gamma^{\widetilde {\rm MOM}}_2 / \alpha(\mu)$ over the whole window. From the previous works~\cite{propag,luscher} and 
from the present study we see that the non-asymptoticity has a tendency to provoke overestimates of $\Lambda^{\overline {\rm MS}}$ (the effective $\Lambda^{\overline {\rm MS}}_{\rm (3 loop)}$ decreases as the matching is performed at higher energies). This observation is sufficient to exclude the negative values of $\widehat{\gamma_3}$. Then, by varying $0 \leq \widehat{\gamma_3} \le 4 \pi 
\gamma^{\widetilde {\rm MOM}}_2/ \alpha(5.6  \rm{GeV}) \sim 42000$
we obtain the uncertainty in $\Lambda^{\overline {\rm MS}}$ of $\sim 20$ MeV. 
Thus, the Landau
gauge gluon propagator analysis results in

\beq
\Lambda^{\overline {\rm MS}}=319 \pm 14 ^{+10}_{-20}  \ ,\label{319}
\eeq

\noindent where the upper limit for the systematic uncertainty comes from the dispersion we 
observed by exploring the domain of {\it three-loop good schemes}, around ${\widetilde {\rm MOM}}$, 
over the ($\gamma_1,\gamma_2$)-plane. It is important to note that the 
uncertainty on the value of 
$\Lambda^{\overline {\rm MS}}$ would be considerably reduced if the value
 $\gamma^{\widetilde {\rm MOM}}_3$ was known. 
 Not only the $-20$~MeV would be reduced, but also 
 the dispersion over the set of  {\it good schemes} to four loops would 
 be fairly restrained.
 
 Reciprocally, if $\Lambda^{\overline {\rm MS}}$ was known accurately from any other source, we could use
 our data to fit $\widehat{\gamma}^3$ rather accurately. For example taking $\Lambda^{\overline {\rm MS}}$
 strictly equal to the central value in eq. (\ref{319}), would give 
 
 \beq
 \left .\frac{\gamma_3^{\widetilde{\rm{MOM}}}}{(4\pi)^4} + \frac{\beta_3^{\widetilde{\rm{MOM}}}}{(4\pi)^4} 
\frac{\gamma_0}{\beta_0}\quad \right|_{\Lambda^{\overline {\rm MS}}=319\, {\rm MeV}} =\ 0.88 \pm  0.04.
\label{hat3}
 \eeq
 where we have used eq. (\ref{condition}). But let us repeat, the fitted value for $\widehat{\gamma}^3$
 varies quickly when $\Lambda^{\overline {\rm MS}}$ is varied within the error bars, which explains 
 the large error in (\ref{resZb}) or equivalently:
 \beq
 \frac{\gamma_3^{\widetilde{\rm{MOM}}}}{(4\pi)^4} + \frac{\beta_3^{\widetilde{\rm{MOM}}}}{(4\pi)^4} 
\frac{\gamma_0}{\beta_0} = 0.9 \pm 0.6.
 \eeq

%%%%%%%%%%%%%%%%%%%%%%%%%%%%%%%%%%%%%%%%%%
%%%%%%%%%%%%%%%%%%%%%%%%%%%%%%%%%%%%%%%%%%
%%%%%%%%%%%%%%%%%%%%%%%%%%%%%%%%%%%%%%%%%%
%%%%%%%%%%%%%%%%%%%%%%%%%%%%%%%%%%%%%%%%%%
\section{Discussion}
\label{discussion}
%%%%%%%%%%%%%%%%%%%%%%%%%%%%%%%%%%%%%%%%%%
%%%%%%%%%%%%%%%%%%%%%%%%%%%%%%%%%%%%%%%%%%
%%%%%%%%%%%%%%%%%%%%%%%%%%%%%%%%%%%%%%%%%%

The estimate of $\Lambda^{\overline{{\rm MS}}}$ is significantly higher than the one obtained
with the Schr\"odinger functional~\cite{luscher}, 
$\Lambda^{\overline{{\rm MS}}}=238\pm 19$ MeV. Zero flavour NRQCD results~\cite{davies}, 
although not directly expressed in terms of $\Lambda^{\overline{{\rm MS}}}$, seem to agree with
the result from Schr\"odinger functional. Estimates from string tension cover a large range of values: 
$244(8)$~MeV~\cite{bali}, $293(18)^{+25}_{-63}$~MeV~\cite{bali2}.
On the other hand, the value recently obtained directly from the triple 
gluon vertex, $\Lambda^{\overline{{\rm MS}}}=295\pm 20$ MeV in~\cite{alpha}, and the less recent 
$340(50)$~MeV~\cite{alles}, along with the one obtained in this paper, favor the larger 
values of $\Lambda^{\overline{{\rm MS}}}$. 
The discrepancy is of the order of three sigma. Method based on Schr\"odinger functional and the one
 based on Green functions are quite different so that a direct comparison is not easy. 
 Could it be that the reason for this discrepancy is simply that we did not 
 reach a enough large energy~?
 In other words, could it be that the difference between  $\Lambda^{\overline{{\rm MS}}}= 238 \pm 19$ MeV
 and the result (\ref{319}) were simply due to the fact that next to third order terms, which are not used in
 the fit leading to (\ref{319}), do mimic a larger $\Lambda^{\overline{{\rm MS}}}$~?
   To investigate this question we 
 use a simple check: had we assumed the result for $\Lambda^{\overline{{\rm MS}}}$ in 
 ref.~\cite{luscher} to be the right asymptotic one, we would have obtained
\beq \left .
\frac{\gamma_3^{\widetilde{\rm{MOM}}}}{(4\pi)^4} + \frac{\beta_3^{\widetilde{\rm{MOM}}}}{(4\pi)^4} 
\frac{\gamma_0}{\beta_0}\quad \right|_{\Lambda^{\overline {\rm MS}}=238 \,{\rm MeV}} =  \ 8.42 \pm 0.08 \ , 
\label{hat3lusch}
\eeq
\noindent over the same energy window used for~(\ref{hat3}), $(2.8\div 9.5)$~GeV, the $\chi^2/{\rm d.o.f.}$ 
being $0.89$. The Schr\"odinger functional result
applied to our data would then imply that the Landau gauge gluon propagator is not asymptotic 
at three-loops at the energy scale of $9$~GeV. The four-loop contribution in this case would be
 much bigger than the three-loop one (about four times). This seems rather unlikely!
We therefore conclude that the value $\Lambda^{\overline{{\rm MS}}}$ obtained by using the 
Schr\"odinger functional technique is difficult to accommodate with the gluon propagator data by 
using the three-loop expression.
There may be some unknown systematic effect explaining this discrepancy.

Let us finally make a comment about the convergence of the gluon propagator. 
The direct connection between the renormalization constant in 
the ${\widetilde {\rm MOM}}$-scheme and the gluon 
propagator makes the gluon momentum the natural scale in this scheme. The scales in the 
${\widetilde {\rm MOM}}$-scheme are significantly larger than the corresponding ones in the 
${\overline {\rm MS}}$ scheme, typically by a factor 1/0.346 (see eq. (\ref{lambdamom}).
The Landau pole in the ${\widetilde {\rm MOM}}$ scheme is around $1$~GeV. 
As  the perturbative regime of the propagator is expected to settle in 
when $\log(\mu/\Lambda_{{\widetilde {\rm MOM}}})$ is large enough,  
it is then not too surprising that high perturbative orders are important around $\sim 3$~GeV 
and higher energies are needed to reach a three-loop perturbative scaling.

%%%%%%%%%%%%%%%%%%%%%%%%%%%%%%%%%%%%%%%%%%
%%%%%%%%%%%%%%%%%%%%%%%%%%%%%%%%%%%%%%%%%%
%%%%%%%%%%%%%%%%%%%%%%%%%%%%%%%%%%%%%%%%%%
%%%%%%%%%%%%%%%%%%%%%%%%%%%%%%%%%%%%%%%%%%
\section{Conclusions}
%%%%%%%%%%%%%%%%%%%%%%%%%%%%%%%%%%%%%%%%%%
%%%%%%%%%%%%%%%%%%%%%%%%%%%%%%%%%%%%%%%%%%
%%%%%%%%%%%%%%%%%%%%%%%%%%%%%%%%%%%%%%%%%%

The main goal of the present work was to go deeper into 
the study of the asymptoticity of the Landau gauge gluon propagator. The matching of its 
non-perturbative evaluation from lattice with perturbative predictions, gives us an 
estimate of the strong coupling constant and hence of $\Lambda^{\rm{QCD}}$~\cite{propag}. 

We have carefully examined the lattice spacing effects, particularly
the hypercubic artifacts, and the finite volume ones. 
We find a close linearity in $a^2 p^{[4]}$ of the gluon propagator, with the slope given 
by eq.~(\ref{pentes}) which 
removes efficiently the hypercubic artifacts. 
The finite volume effects, in the region of large momenta, are 
parametrized by the relation~(\ref{vol}). 

After having subtracted the lattice artifacts,
we found that the four-loop contribution is negligible above $\mathsf{\sim 5.}$~GeV, but 
 becomes important below this energy, confirming the conclusion of ref.~\cite{propag}.
 In its turn the four-loop
perturbative scaling fails below $\mathsf{2.8}$ GeV: 
the Landau gauge gluon propagator reaches very slowly the asymptotia.

We therefore have fitted with a 
three-loop formula over the 
energy window $5.6\ {\rm GeV}\  \leq \  \mu \ \leq \ 9.5\ {\rm GeV}$. 
The rather good fit leads to 
$\Lambda_{(3 \rm{loop})}^{\overline{{\rm MS}}} \ = \ 319 \pm 14$~MeV.
 A fitted four-loop formula has been used to extend the fit over the larger energy window 
 ($2.8 \div 9.5$)~GeV. 
We have obtained a consistent description of all our lattice data. 

 Our final result is
\beq
\Lambda^{\overline{{\rm MS}}} \ = \ 319 \pm 14 ^{+10}_{-20} \; \; \; 
   \frac{a^{-1}(6.2)}{2.75 {\rm GeV}}  \; \; \;{\rm MeV} \ \ ,
\eeq
with the errors discussed in detail in Sec.~\ref{pattern}.  
Although a combination of theoretical results is always delicate,
we may try to combine this result with the one obtained from the study of the three 
gluon vertex~\cite{alpha}, $\Lambda^{\overline{{\rm MS}}} = 295\pm 20$~MeV. This results 
in an overall flavorless estimate from the gluon Green functions : $
275 \ \rm{MeV} \leq \Lambda^{\overline{{\rm MS}}} \leq 343 \ \rm{MeV} $ .

\section*{Acknowledgments.}

These calculations were performed on the QUADRICS QH1 located in the Centre de Ressources
 Informatiques (Paris-sud, Orsay) and purchased thanks to a funding from the Minist\`ere de
  l'Education Nationale and the CNRS. D.B. acknowledges the Italian INFN, and J.R.Q. the 
  Spanish Fundaci\'{o}n Ram\'{o}n Areces for financial support. 

%%%%%%%%%%%%%%%%%%%%%%%%%%%%%%%%%%%%%%%%%%%%%%%%%%%%%%%
%%%%%%%%%%%%%%%%%%%%%%%%%%%%%%%%%%%%%%%%%%%%%%%%%%%%%%%

\end{document}